# Hidden Markov Models on Variable Blocks with a Modal Clustering Algorithm and Applications

Lin Lin[*]    Jia Li[†]

June 28, 2016

**Abstract**

Motivated by high-throughput single-cell cytometry data with applications to vaccine development and immunological research, we consider statistical clustering in large-scale data that contain multiple rare clusters. We propose a new hierarchical mixture model, namely Hidden Markov Model on Variable Blocks (HMM-VB), and a new mode search algorithm called Modal Baum-Welch (MBW) for efficient clustering. Exploiting the widely accepted chain-like dependence among groups of variables in the cytometry data, we propose to treat the hierarchy of variable groups as a figurative time line and employ a HMM-type model, namely HMM-VB. We also propose to use mode-based clustering, aka modal clustering, and overcome the exponential computational complexity by MBW. In a series of experiments on simulated data HMM-VB and MBW have better performance than existing methods. We also apply our method to identify rare cell subsets in cytometry data and examine its strengths and limitations.

[*][†]Department of Statistics, Pennsylvania State University, University Park, PA 16802, USA; e-mail: {llin,jiali}@psu.edu

0

# 1. Introduction

Multivariate mixture model is arguably the most widely used statistical method for clusters identification. In this paper, we develop a novel hierarchical mixture modeling approach aimed at large-scale moderately high dimensional data sets that contain multiple rare clusters. The work is motivated by single-cell studies with applications to vaccine development and immunological research. The conventional mixture model approach is not up to the challenge of capturing clusters with highly unbalanced sizes even though the dimensionality of such data is in the tens. In order to model the data accurately without missing the rare clusters, a large number of mixture components is needed. On the other hand, the curse of dimensionality prevents the use of many components. Furthermore, the increase of components results in higher computational complexity. To resolve this dilemma, we leverage a natural chain-like dependence among groups of variables in such data, a key fact that biologists and clinicians have been utilizing in exploratory studies. Existing statistical modeling techniques do not exploit this important characteristic. Our method is an effort to bridge the gap.

## 1.1 Background on Single-cell Cytometry Data Analysis

Interrogation of cell population heterogeneity has been made possible by the recent advancements in single-cell cytometry technologies (Perfetto et al. 2004; Bandura et al. 2009; Maecker et al. 2012; Chattopadhyay et al. 2014; Spitzer and Nolan 2016). For example, current high-throughput flow cytometry experiments measure $10 \sim 20$ parameters (cell markers/variables) routinely, including both phenotypic and functional markers, on a large number of single cells (hundreds of thousands to several millions). The current mass cytometry (CyTOF) can now measure up to $50$ parameters at a single cell level.

A key first step to analyze this wealth of data is to partition the data (cells) from a sample (typically blood or tissue) into clusters based on the measured cell markers. The identified clusters are usually referred to as (cell) subsets. In most studies, the sample sizes are large, reaching to several millions. In addition, cell subsets of interest are typically in low frequencies (e.g., $\sim$ 0.01% of total cells). There is a need for detecting cell heterogeneity, especially very low frequency cell subsets for subsequent analysis. For example, association studies which help understand the association between cellular heterogeneity and disease progression (Ciuffreda et al. 2008; Lin et al. 2015a; Seshadri et al. 2015; Corey et al. 2015).

Traditionally, the cell subsets are identified by a manual gating strategy. Specifically, a manual sequential process that visually demarcates cells in bounded regions (called gates) on histogram or 2-D scatter plot projections. Figure 1 provides a simple illustration for the manual gating analysis on one flow cytometry data. Red lines are the gates, and cells within the region defined by the gates are identified as a specific cell subset. For example, to discriminate CD4+ T cells, which is one cell subset, a sequence of subsetting procedures is performed. Two physical markers, Forward (FSC-A) and side (SSC-A) light scatter, are first used to distinguish lymphocytes from all the live cells. Lymphocytes can then be further partitioned based on 3 fluorescence parameters: CD3, CD4 and CD8 cell-surface markers. CD4+ T cells are the subclass of lymphocytes having high values of CD3 and CD4 but low value of CD8. Within CD4+ T-cell populations, additional functional markers such as intracellular makers (IL2 and IFNg) can further distinguish many functionally different CD4+ T-cell subsets. The sequence of markers to use is called *gating hierarchy*, which is determined by expert knowledge. The exact shape and location of the gates are manually drawn. The gates are typically used to dichotomize the continuous marker expressions into binary value: positive and negative. Therefore, the manual gating analysis is subjective and hard to reproduce.

Figure 1: A simple example of cell subsets identification by manual gating analysis. The flow cytometry measurements on single cells from a blood sample are shown using 4 heat maps of 2-D scatter plots projected on different dimensions (markers). The red lines on each subplots are called gates. Cells within the red lines are the subset of interest, which is subsetted and projected on the next subplot in the sequence. The percentages are the frequencies of the identified cell subsets relative to the total number of cells.

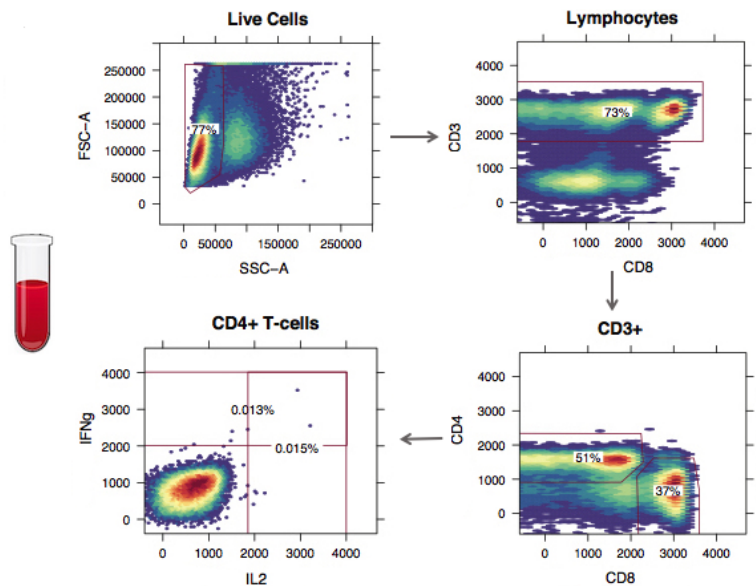

It is evident that the manual gating analysis is essentially a hierarchical clustering procedure. The hierarchy is formed on a chain of variable groups. More specifically, consider the sequential visualization process, any move one step further means a new group of variables are examined and used to select a more refined subset of data from an existing set chosen based on the previous groups of variables. By using the well-established gating hierarchy that projects cells on lower dimensions, each subplot provides a finer resolution of the cellular heterogeneity. In other words, the variables are grouped and ordered based on expert knowledge, and then the data are zoomed in by progressively putting constraints on more and more variable groups.

## 1.2 Automated Clustering Using Mixture Models

To overcome the limitations of the manual gating strategy, probabilistic clustering based on statistical mixture models (Boedigheimer and Ferbas 2008; Lo et al. 2008; Chan et al. 2008; Pyne et al. 2009; Aghaeepour et al. 2013; Lin et al. 2016) have been established to identify cell subsets objectively and automatically. One advantage of the mixture model is that the goodness of fit to any data sample can be improved by increasing the number of mixture components. The simplest approach to identify cell subsets based on a mixture model is to assign each mixture component to an individual cell subset. However, cell subsets can have any arbitrary shape. The parametric distribution of each mixture component is often inadequate to capture different shapes of the subsets. Various strategies have been proposed to merge multiple mixture components so that an individual cell subset can be better modeled (Pyne et al. 2009; Finak et al. 2009; Chan et al. 2010; Aghaeepour et al. 2011; Lin et al. 2016). In the statistical learning literature, the methodology of merging multiple mixture components into one cluster based on the modes of mixture density is called modal clustering (Li et al. 2007).

Although the typical mixture modeling approach has achieved some success, it is not powerful enough to detect clusters of very low probabilities. This is because such low probability mixture components are "consumed" by large background clusters in the data. One may naturally ask why not to increase the number of mixture components dramatically. However, to robustly estimate a mixture model, the number of components is limited by some maximum tolerable value depending on the data size (again curse of dimensionality). Our observation indicates that even the maximum tolerable value is not adequate for detecting those low frequency clusters. Moreover, a practical issue with using more components is the computational burden. It is thus imperative to design and implement effective and computationally accessible novel statistical models that can flexibly and robustly fit data on both high and low probability regions, in addition to the capability of handling non-Gaussian shape subsets.

Comparing the existing statistical mixture modeling approaches with the manual gating strategy, it is not hard to notice that the chain-like dependence among groups of variable (aka gating hierarchy), vital for the feasibility of the manual gating strategy, is not exploited by the aforemen-tioned mixture models. Clearly, if the dependence can be considered, more effective modeling of high dimensional data is possible. The recent work of Lin et al. (2013), which develops a hierarchical nonparametric Bayesian mixture model, explores along this direction. We see it as a prelude of our current work. The key idea of that paper is to partition cell features into two groups, also called subvectors. The first subvector is modeled by a Gaussian mixture model. Given the latent component identity of the first subvector, the second subvector is conditionally independent from the first subvector, and follows another Gaussian mixture distribution. In the conditional mixture distribution of the second subvector, the prior probabilities of the components depend on the latent component identity of the first subvector, while the parameters of individual normal components do not. This modeling feature is achieved via the use of the hierarchical Dirichlet process prior. The automatic clustering mechanism resembles the zoom-in on data in a hierarchical way by manual gating. The latent component identity of the first subvector determines the first level clustering. Clusters obtained from the first level are then further divided based on the latent component identity of the second subvector. It is found that this two-subvector model is more efficient and accurate in rare subset identification than the traditional generative mixture models.

## 1.3 Our Contributions

The next logical extension of Lin et al. (2013) is to accommodate more than two groups of variables, similarly as in the manual gating where a sequence of variable groups are visited. Albeit obvious in concept, there are technical hurdles seemingly impossible to conquer by the approach of hierarchical nonparametric Bayesian. First, when the number of variable groups grows, the computational complexity for estimating the latent component identities grows exponentially. The Markov chain Monte Carlo (MCMC) simulation for the two-subvector model is already highly intensive. There is also the classic issue of label switching when MCMC is applied to estimate the mixture models (e.g., Richardson and Green (1997); Celeux et al. (2000); Stephens (2000)). Practically, the existing method is already hitting a wall just to go beyond two subvectors. Second, a more subtle issue is the exponential growth of combinations of the latent component identities. Suppose there are $k$ variable groups and each group is modeled by a mixture of $M$ components. Then there are $M^k$ combinations of components over all the variables, each can correspond to one cluster. This exponential increase in the number of clusters is unrealistic, raising doubt on the legitimacy of the result. Even if we perform modal clustering, which could dramatically reduce the number of clusters, such algorithm to find the modes again has an exponential complexity.

In this paper, we develop a general statistical model, namely Hidden Markov Model on Variable Blocks (HMM-VB), and overcome both of the aforementioned difficulties by deriving new algorithms for model estimation and mode identification. Although our new model is not a HMM in the conventional sense, a point elaborated in Section 3, the latent graphical dependence structure for the variables is essentially the same as that of the sequential vector data in HMM. We thus put our model in the light of HMM and leverage the powerful computational tools for HMM. To the best of our knowledge, our work is the first to exploit the hierarchical dependence structure among variable groups by HMM.

The estimation algorithm for HMM can resolve the estimation difficulty encountered by the MCMC approach. One natural idea to use HMM-VB to cluster data is to find the underlying sequence of states with maximum posterior. However, the number of possible state sequences is exponential in the sequence length, in practice, often much larger than the data size. To tackle this issue, we cast the HMM-VB back as a Gaussian mixture model and use modal clustering to merge the components in Gaussian mixture model. However, the number of components in the equivalent Gaussian mixture model equals the number of possible state sequences. As a result, a straightforward application of a mode-seeking algorithm, such as Modal EM (MEM) (Li et al. 2007), still faces exponential computational complexity. We discover a new numerical algorithm with linear complexity for finding modes of a HMM-VB, which is proven to be equivalent to MEM. This is one key for the success of clustering based on HMM-VB.

Our experiments on both simulated and real data sets show that the approach of HMM-VB is effective in discovering rare clusters and is robust to non-Gaussian shaped clusters. It naturally incorporates domain knowledge on the dependence structure of the variables. We also explore the potential of HMM-VB as a general tool for parsimonious mixture modeling. In particular, we test HMM-VB on data not complying to the chain-like dependence of variable blocks and comparisons with the general mixture models are performed.

The rest of the paper is organized as follows. In Section 2, we introduce notations and overview existing techniques most relevant to our proposed methods. In Section 3, we describe the proposed model and develop efficient algorithms for model fitting and for performing modal clustering. Section 4 presents the results of both simulation studies and the application to large-scale cytometry data. Comparisons are made with some competing models. Finally, we summarize and discuss our contributions in Section 5.

## 2. Preliminaries

The finite Gaussian mixture model (GMM) is commonly used for clustering. For random variable $X \in \mathcal{R}^d$, where $d$ is the number of dimensions, a GMM with $M$ components has density function:

$$f(x|\theta) = \sum_{k=1}^{M} \pi_k \phi(x|\theta_k), \tag{1}$$

where $\pi_k$ is the mixture component prior probability and $\phi(\cdot \mid \theta_k)$ is the multivariate normal density parameterized by $\theta_k = (\mu_k, \Sigma_k)$, $\mu_k$ being the $d$-dimensional mean vector and $\Sigma_k$ the $d \times d$ covariance matrix. The collection of parameters $\theta = (\pi_1, ..., \pi_M, \theta_1, ..., \theta_M)$.

For model estimation, a latent indicator $Z \in \{1, 2, ..., M\}$ with $P(Z = k) = \pi_k$ is used. Specifically, $Z = k$ if, and only if, $X$ comes from component $k$. $Z$ is also called the component identity of $X$. To perform clustering, the usual approach is to compute the posterior probability $P(Z = k \mid X = x)$, and assign $X$ to the cluster $k$ with the maximum posterior. However, this approach is inadequate to model clusters with arbitrary shapes and cannot ensure that the clusters are reasonably separated. One major idea explored in the literature is to merge multiple mixture components for a better and more flexible representation of an individual cluster. We refer to (Melnykov and Maitra 2010) for a thorough review on clustering based on finite mixture models.

### 2.1 MEM for Gaussian Mixtures and Modal Clustering

The Modal EM (MEM) algorithm developed by Li et al. (2007) performs efficient merging of mixture components. It resembles the expectation-maximization (EM) algorithm (Dempster et al. 1977), as reflected by the name "modal EM". However, the optimization objective of MEM is the local maxima over $x$ for $f(x \mid \theta)$ under a given $\theta$, while EM is to find local maxima over $\theta$ for $f(x \mid \theta)$ given $x$. Consider a general mixture density $f(x) = \sum_{k=1}^{M} \pi_k f_k(x)$, where $f_k(x)$ is the density of the $k$th mixture component. Staring from any initial value $x^{[0]}$, MEM generates a local maximum of the mixture density by performing the following two iterative steps: (1) At iteration $r$, let $p_k = \frac{\pi_k f_k(x^{[r]})}{f(x^{[r]})}$, $k = 1, ..., M$; (2) Update $x^{(r+1)} = \mathrm{argmax}_x \sum_{k=1}^{M} p_k \log f_k(x)$. MEM stops when a pre-specified stopping criterion is met. Specifically for GMM $f(x) = \sum_{k=1}^{M} \pi_k \phi(x|\mu_k, \Sigma_k)$, MEM becomes

1. E-step: solve

$$p_k = \frac{\pi_k \phi_k(x^{[r]} \mid \mu_k, \Sigma_k)}{f(x^{[r]})}, \quad k = 1, ..., M. \tag{2}$$

2. M-step: solve

$$x^{[r+1]} = \left(\sum_{k=1}^{M} p_k \cdot \Sigma_k^{-1}\right)^{-1} \cdot \left(\sum_{k=1}^{M} p_k \cdot \Sigma_k^{-1} \mu_k\right). \tag{3}$$

The computational efficiency of MEM enabled the development of a new clustering approach by Li et al. (2007), referred to as *modal clustering*. In Li et al. (2007), a nonparametric Gaussian kernel density estimate is used; and initialized by each data point, MEM is applied to find the mode associated with every point. Data points associated to the same mode are classified to be in the same cluster. In Lee and Li (2012), a natural extension is developed for the general GMM. To save computation, instead of applying MEM to every data point, it is applied to the means of the mixture components. Components (hence data points) with mean vectors associated to the same mode are merged into one cluster. Whether a point-wise mode association or a component-wise mode association is preferred depends on the nature of the application and the computational resource. In practice, the difference in the clustering results we have observed is quite small. For the finite mixture model, the number of components is usually much smaller than the data size. Hence component-wise mode association saves computation. In some situation, the number of mixture components can be enormous (much more than the data size), as is the case for our hierarchical mixture model to be introduced in Section 3. Clearly, a point-wise mode association would be preferred. However, finding a mode starting from every data point can still be costly for big data. As a result, we take the scheme of component-wise mode search, but we do not find the associated mode for every component mean. Instead, we first identify all the components that have been chosen by at least one data point according to the maximum a posteriori criterion. The number of such components is upper bounded by the data size and is often much smaller. The modes are then computed with initialization set by each of these chosen component means.

Under the framework of modal clustering, the role of a mixture component is primarily for achieving good density estimation. We no longer rely on a component to represent a cluster. This provides flexibility in modeling. We can allow components to overlap substantially for the sake of more accurate modeling of the density. When it comes to partitioning data, mode association ensures that different groups of data are sufficiently separated. On a practical side, the choice of the number of components is not as crucial as the conventional mixture model based clustering, where that number determines how many clusters are generated. In fact, in the modal clustering algorithm in Li et al. (2007), the Gaussian kernel density is used, where the number of components equals the data size. More elaborated discussions on the advantage of modal clustering and comparisons with the usual mixture model clustering are provided in that paper. Interestingly, the modal clustering technique has been applied to cytometric data analysis by Ray and Pyne (2012). A computational framework called flowScape was developed to mimic the analytical actions of human experts in the process of manual gating (Ray and Pyne 2012). The purpose of that work is to enhance the manual gating process, while our work belongs to the school of automated clustering.

The flexibility provided by modal clustering for modeling data is precisely what we need for the applications we consider. As will be explained in Section 3, we exploit a HMM-type model which can be cast as a mixture model with an enormous number of components, even exceeding the data size. This causes no difficulty in clustering via mode association.

## 2.2 Hidden Markov Model

Consider sequential data $\mathbf{x} = \{x_1, x_2, ..., x_t, ..., x_T\}$, $x_t \in \mathcal{R}^d$. As in the mixture model, assume there is an underlying state $s_t$ associated with every $x_t$, $t = 1, ..., T$. The underlying state is the counterpart of the mixture component identity in GMM. The state $s_t \in \mathcal{S} = \{1, 2, ..., M\}$, where $M$ is the number of states. Let the set of all possible state sequences be $\bar{\mathcal{S}}$, that is, the set of $T$-tuples on $\mathcal{S}$. $|\bar{\mathcal{S}}| = M^T$. Denote $\mathbf{s} = \{s_1, ..., s_T\} \in \bar{\mathcal{S}}$. The basic assumptions of a HMM are:

1. The underlying states $\{s_1, s_2, ..., s_T\}$ follow a Markov chain. The Markov chain is usually time invariant with transition probability matrix $A = (a_{k,l})$, where $a_{k,l} = P(s_{t+1} = l \mid s_t = k)$, $k, l \in \mathcal{S}$. Let the initial probabilities of states be denoted by $\pi_k = P(s_1 = k)$, $k \in \mathcal{S}$.

2. Given the hidden state $s_t$, the observation $x_t$ is conditionally independent from $s_{t'}$ and $x_{t'}$ for any $t' \neq t$; and the distribution of $x_t$ given $s_t$ depends on $s_t$, but not $t$. Denote the conditional the density of $P(x_t = x|s_t = k)$ by $b_k(x)$. In particular, $b_k(x) = \phi(x|\mu_k, \Sigma_k)$.

In summary:

$$P(\mathbf{x}, \mathbf{s}) = P(\mathbf{s})P(\mathbf{x} \mid \mathbf{s}) = \pi_{s_1} b_{s_1}(x_1) a_{s_1, s_2} b_{s_2}(x_2) \cdots a_{s_{T-1}, s_T} b_{s_T}(x_T) ,$$
$$P(\mathbf{x}) = \sum_{\mathbf{s} \in \bar{\mathcal{S}}} P(\mathbf{s})P(\mathbf{x} \mid \mathbf{s}) = \sum_{\mathbf{s} \in \bar{\mathcal{S}}} \pi_{s_1} b_{s_1}(x_1) a_{s_1, s_2} b_{s_2}(x_2) \cdots a_{s_{T-1}, s_T} b_{s_T}(x_T).$$

The parameters to be estimated in a HMM are the transition probabilities: $a_{k,l}$, $k, l = 1, ..., M$, the initial probabilities: $\pi_k$, $k = 1, ..., M$, and $\mu_k$, $\Sigma_k$ for each state $k = 1, ..., M$. HMM is usually estimated by the EM algorithm. However, because the cardinality of $\bar{\mathcal{S}}$ grows exponentially with the sequence length, the computational complexity of a direct application of EM is of exponential complexity. This technical hurdle was overcome by the *Baum-Welch (BW) algorithm* that achieves without suffering optimality a complexity linear in the sequence length and quadratic in the number of states. The BW algorithm, a special instance of EM, was developed in the 1960's before the general EM algorithm was developed in the 1970's. As a result, we still call the estimation algorithm Baum-Welch, following the convention of the literature on HMM.

Under a set of parameters, let $L_k(t)$ be the conditional probability of being in state $k$ at position $t$ given the entire observed sequence $\mathbf{x} = \{x_1, x_2, ..., x_T\}$. Let $I(\cdot)$ be the indicator function that equals $1$ when the argument is true and $0$ otherwise. Then

$$L_k(t) = P(s_t = k|\mathbf{x}) = \sum_{\mathbf{s}} P(\mathbf{s} \mid \mathbf{x}) I(s_t = k), \quad k \in \mathcal{S}. \tag{4}$$

Let $H_{k,l}(t)$ be the conditional probability of being in state $k$ at position $t$ and being in state $l$ at position $t+1$, i.e., seeing a transition from $k$ to $l$ at $t$, given the entire observed sequence $\mathbf{x}$.

$$\begin{aligned} H_{k,l}(t) &= P(s_t = k, s_{t+1} = l|\mathbf{x}) \\ &= \sum_{\mathbf{s}} P(\mathbf{s} \mid \mathbf{x}) I(s_t = k) I(s_{t+1} = l), \quad k, l \in \mathcal{S}. \end{aligned} \tag{5}$$

Note that $L_k(t) = \sum_{l=1}^{M} H_{k,l}(t)$, $\sum_{k=1}^{M} L_k(t) = 1$. Since $\mathbf{s} \in \bar{\mathcal{S}}$ and the $|\bar{\mathcal{S}}| = M^T$, it is infeasible to compute $L_k(t)$ and $H_{k,l}(t)$ by the above equations directly. As part of the BW algorithm, the *forward-backward* algorithm is used to compute $L_k(t)$ and $H_{k,l}(t)$ efficiently. The amount of computation needed is at the order of $M^2 T$; and memory required is at the order of $MT$. We present the forward-backward algorithm in Appendix A. For the current discussion, suppose the quantities have been solved.

The BW algorithm iterates the following two steps:

1. E-step: Under the current set of parameters, compute $L_k(t)$ and $H_{k,l}(t)$, for $k, l = 1, ..., M$, $t = 1, ..., T$.

2. M-step: Update parameters by

$$\mu_k = \frac{\sum_{t=1}^{T} L_k(t) x_t}{\sum_{t=1}^{T} L_k(t)}, \quad \Sigma_k = \frac{\sum_{t=1}^{T} L_k(t)(x_t - \mu_k)(x_t - \mu_k)^t}{\sum_{t=1}^{T} L_k(t)}, \quad a_{k,l} = \frac{\sum_{t=1}^{T-1} H_{k,l}(t)}{\sum_{t=1}^{T-1} L_k(t)}.$$

The initial probabilities of states $\pi_k$ are often manually determined. We can also estimate them by $\pi_k \propto \sum_{t=1}^{T} L_k(t)$, subject to $\sum_{k=1}^{M} \pi_k = 1$, or $\pi_k \propto L_k(1)$.

So far we present the BW algorithm in the case of estimation based on a single sequence. The extension to estimation from multiple sequences is relatively easy, which we describe in Appendix A. In our work, as will be explained in the next section, we always handle multiple sequences. For a detailed exposure to HMM, we refer to (Young et al. 1997).

# 3. Hidden Markov Model on Variable Blocks

Consider $X \in \mathcal{R}^d$. We propose a dependence structure among groups of variables that in essence matches the statistical dependence among observations at different time spots assumed by HMM. We call the model *Hidden Markov Model on Variable Blocks (HMM-VB)*. Suppose the variables are divided into blocks $t = 1, 2, ..., T$, where $T$ is the total number of blocks. Let the number of variables in block $t$ be $d_t$, $d = \sum_{t=1}^{T} d_t$. In a nutshell, HMM-VB treats the block index $t$ as "time". The variables in the same block are treated as multivariate data and their joint density is assumed to be a GMM. The Gaussian component identity of a variable block acts as the underlying state for this block. To capture the dependence between the variable blocks, the states are assumed to follow a Markov chain. As discussed in Section 1.3, such a dependence structure is motivated by the gating hierarchy among groups of variables, which has been well accepted by researchers on cytometry data analysis.

Without loss of generality, suppose we have ordered the variables in a way such that the first $d_1$ variables are in block 1, the next $d_2$ variables are in block 2, and so on. Let *variable blocks* $x^{(1)} = (x_1, x_2, ..., x_{d_1})'$ and $x^{(t)} = (x_{m_t+1}, x_{m_t+2}, ..., x_{m_t+d_t})'$, where $m_t = \sum_{\tau=1}^{t-1} d_\tau$, $t = 2, ..., T$. Denote the underlying state for $x^{(t)}$ by $s_t$, $t = 1, ..., T$. Let the index set of $s_t$ be $\mathcal{S}_t = \{1, 2, ..., M_t\}$, where $M_t$ is the number of mixture components for variable block $x^{(t)}$, $t = 1, ..., T$. Let the set of all possible sequences be $\hat{\mathcal{S}} = \mathcal{S}_1 \times \mathcal{S}_2 \cdots \times \mathcal{S}_T$. $|\hat{\mathcal{S}}| = \prod_{t=1}^{T} M_t$. We assume:

1. $\{s_1, s_2, ..., s_T\}$ follow a Markov chain. Let $\pi_k = P(s_1 = k)$, $k \in \mathcal{S}_1$. Let the transition probability matrix $A_t = (a_{k,l}^{(t)})$ between $s_t$ and $s_{t+1}$ be defined by $a_{k,l}^{(t)} = P(s_{t+1} = l | s_t = k)$, $k \in \mathcal{S}_t$, $l \in \mathcal{S}_{t+1}$.

2. Given $s_t$, $x^{(t)}$ is conditionally independent from the other $s_{t'}$ and $x^{(t')}$, $t' \neq t$. We also assume that given $s_t = k$, the conditional density of $x^{(t)}$ is the Gaussian distribution $\phi(x^{(t)} \mid \mu_k^{(t)}, \Sigma_k^{(t)})$.

Denote $\mathbf{s} = \{s_1, ..., s_T\}$. To summarize, the density of HMM-VB is given by

$$f(x) = \sum_{\mathbf{s} \in \hat{\mathcal{S}}} \left( \pi_{s_1} \prod_{t=1}^{T-1} a_{s_t, s_{t+1}}^{(t)} \right) \cdot \prod_{t=1}^{T} \phi(x^{(t)} | \mu_{s_t}^{(t)}, \Sigma_{s_t}^{(t)}) . \tag{6}$$

Remarks on comparison with the conventional HMM:

1. The variable blocks $x^{(t)}$'s are not from the same vector space. Hence, the parameters of the distribution of $x^{(t)}$ given $s_t = k$ depend not only on $k$ but also on $t$.

2. The underlying Markov chain for $\{s_1, ..., s_T\}$ is not time invariant. In fact, the the state space $\mathcal{S}_t$ varies with $t$.

## 3.1 Maximum Likelihood Estimation

We derive the corresponding BW algorithm for HMM-VB. It is an instance of the EM algorithm, similarly as the BW algorithm for the usual HMM. The proof of the BW algorithm as an equivalence of the EM algorithm for HMM-VB follows closely that for the usual HMM. We thus omit it here. Clearly, it is not meaningful to estimate HMM-VB using a single sequence, aka, a single data point. HMM-VB is after all a model for $X \in \mathcal{R}^d$. The imposed sequence structure is for the sake of modeling dependence among the variables, motivated by the nature of the data encountered in applications.

Let $\mathbf{x} = (x^{(1)}, x^{(2)}, ..., x^{(T)}) \in \mathcal{R}^d$ (assume column-wise concatenation) be the full-dimensional data. Consider estimation of HMM-VB based on a data set $\{\mathbf{x}_1, \mathbf{x}_2, ..., \mathbf{x}_n\}$, $\mathbf{x}_i \in \mathcal{R}^d$, $i = 1, ..., n$. Use $x_i^{(t)} \in \mathcal{R}^{d_t}$ to denote the $t$th variable block of $\mathbf{x}_i$, and the notation $x_i^{(t)} = (x_{i,m_t+1}, x_{i,m_t+2}, ..., x_{i,m_t+d_t})'$ for the variables in $x_i^{(t)}$, where $m_0 = 0$ and $m_t = \sum_{\tau=1}^{t-1} d_\tau$, $\tau = 2, ..., T$. Making the dependence on $\mathbf{x}$ explicit in the notation, we define $L_k(\mathbf{x}, t)$ and $H_{k,l}(\mathbf{x}, t)$ similarly as in Eq. (4) and Eq. (5).

$$L_k(\mathbf{x}, t) = P(s_t = k \mid \mathbf{x}), \ k \in \mathcal{S}_t, \tag{7}$$

$$H_{k,l}(\mathbf{x}, t) = P(s_t = k, s_{t+1} = l \mid \mathbf{x}), \ k \in \mathcal{S}_t, l \in \mathcal{S}_{t+1}. \tag{8}$$

The BW algorithm iterates the following two steps:

1. E-step: Under the current set of parameters, compute $L_k(\mathbf{x}_i, t)$, $i = 1, ..., n$, $k \in \mathcal{S}_t$, $t = 1, ..., T$, and $H_{k,l}(\mathbf{x}_i, t)$, $i = 1, ..., n$, $k \in \mathcal{S}_t$, $l \in \mathcal{S}_{t+1}$, $t = 1, ..., T-1$.

2. M-step: Update parameters by

$$\mu_k^{(t)} = \frac{\sum_{i=1}^n L_k(\mathbf{x}_i, t) x_i^{(t)}}{\sum_{i=1}^n L_k(\mathbf{x}_i, t)}, \quad k \in \mathcal{S}_t, t = 1, ..., T,$$

$$\Sigma_k^{(t)} = \frac{\sum_{i=1}^n L_k(\mathbf{x}_i, t) \left(x_i^{(t)} - \mu_k^{(t)}\right)\left(x_i^{(t)} - \mu_k^{(t)}\right)'}{\sum_{i=1}^n L_k(\mathbf{x}_i, t)}, \quad k \in \mathcal{S}_t, t = 1, ..., T,$$

$$a_{k,l}^{(t)} = \frac{\sum_{i=1}^n H_{k,l}(\mathbf{x}_i, t)}{\sum_{i=1}^n L_k(\mathbf{x}_i, t)}, \quad k \in \mathcal{S}_t, l \in \mathcal{S}_{t+1}, t = 1, ..., T-1,$$

$$\pi_k \propto \sum_{i=1}^n L_k(\mathbf{x}_i, 1), k \in \mathcal{S}_1, \ s.t. \sum_{k \in \mathcal{S}_1} \pi_k = 1.$$

The above equations easily extend to the case of weighted sample points. It can occur in practice that each sample point is assigned with a weight. For instance, quantization is often used to reduce the data size significantly. Instead of the original data, one may use the quantized points, each of which can represent a different number of original points and hence is assigned with a weight proportional to that number. Suppose weight $w_i$ is assigned to sample $\mathbf{x}_i$. The E-step is not affected. In the update of parameters, we can simply multiply $w_i$ in front of each summand appeared in the equations above.

The forward-backward algorithm for computing $L_k(\mathbf{x}_i, t)$ and $H_{k,l}(\mathbf{x}_i, t)$ is essentially the same as the forward-backward algorithm for the usual HMM. The fact that the variable blocks are not from the same vectors space and the state spaces vary with $t$ does not cause any intrinsic difference. Define the forward probability $\alpha_k(\mathbf{x}, t)$ as the joint probability of observing the first $t$ variable blocks $x^{(\tau)}$, $\tau = 1, ..., t$, and being in state $k$ at time $t$:

$$\alpha_k(\mathbf{x}, t) = P(x^{(1)}, x^{(2)}, ..., x^{(t)}, s_t = k), \quad k \in \mathcal{S}_t.$$

This probability can be evaluated by the following recursive formula:

$$\begin{aligned}
\alpha_k(\mathbf{x}, 1) &= \pi_k \phi(x^{(1)} \mid \mu_k^{(1)}, \Sigma_k^{(1)}), \quad k \in \mathcal{S}_1, \\
\alpha_k(\mathbf{x}, t) &= \phi(x^{(t)} \mid \mu_k^{(t)}, \Sigma_k^{(t)}) \sum_{l \in \mathcal{S}_{t-1}} \alpha_l(\mathbf{x}, t-1) a_{l,k}^{(t-1)}, \quad 1 < t \leq T, \; k \in \mathcal{S}_t.
\end{aligned}$$

Define the backward probability $\beta_k(\mathbf{x}, t)$ as the conditional probability of observing the variable blocks after time $t$, $x^{(\tau)}$, $\tau = t+1, ..., T$, given the state at block $t$ is $k$:

$$\begin{aligned}
\beta_k(\mathbf{x}, t) &= P(x^{(t+1)}, ..., x^{(T)} \mid s_t = k), \quad 1 \leq t \leq T-1, \; k \in \mathcal{S}_t, \\
&\text{Set } \beta_k(\mathbf{x}, T) = 1, \quad \text{for all } k \in \mathcal{S}_T.
\end{aligned}$$

The backward probability can be evaluated using the following recursion:

$$\begin{aligned}
\beta_k(\mathbf{x}, T) &= 1, \quad k \in \mathcal{S}_T, \\
\beta_k(\mathbf{x}, t) &= \sum_{l \in \mathcal{S}_{t+1}} a_{k,l}^{(t)} \phi(x^{(t+1)} \mid \mu_l^{(t+1)}, \Sigma_l^{(t+1)}) \beta_l(\mathbf{x}, t+1), \quad 1 \leq t < T, \; k \in \mathcal{S}_t.
\end{aligned}$$

The probabilities $L_k(\mathbf{x}, t)$ and $H_{k,l}(\mathbf{x}, t)$ are solved by

$$L_k(\mathbf{x}, t) = P(s_t = k \mid \mathbf{x}) = \frac{P(\mathbf{x}, s_t = k)}{P(\mathbf{x})} = \frac{\alpha_k(\mathbf{x}, t) \beta_k(\mathbf{x}, t)}{P(\mathbf{x})}, \quad k \in \mathcal{S}_t,$$

$$H_{k,l}(\mathbf{x}, t) = P(s_t = k, s_{t+1} = l \mid \mathbf{x}) = \frac{P(\mathbf{x}, s_t = k, s_{t+1} = l)}{P(\mathbf{x})}$$
$$= \frac{1}{P(\mathbf{x})} \alpha_k(\mathbf{x}, t) a_{k,l}^{(t)} \phi(x^{(t+1)} \mid \mu_l^{(t+1)}, \Sigma_l^{(t+1)}) \beta_l(\mathbf{x}, t+1), \quad k \in \mathcal{S}_t, \ l \in \mathcal{S}_{t+1}.$$

The normalizing factor $P(\mathbf{x}) = \sum_{k \in \mathcal{S}_t} \alpha_k(\mathbf{x}, t) \beta_k(\mathbf{x}, t)$ for any $t$ (equation holds for any $t$).

To initialize the model, we design several schemes. In our experiments, models from different initializations are estimated and the one is chosen with maximum likelihood. In our baseline initialization scheme, k-means clustering is applied individually to each variable block using all the data instances. Based on the clustering result of k-means, we take every cluster as one mixture component and compute the sample mean and sample covariance matrix of data in that cluster. To reduce the sensitivity to the initial clustering result, we also compute the pooled common sample covariance matrix for the clusters. The initial covariance matrix of a component is then set to be a convex combination of the cluster-specific sample covariance and the common sample covariance. The transition probabilities are always initialized to be uniform. Under the second initialization scheme, we randomly sample a subset from the whole data and apply the baseline initialization to the subset. Under the third initialization scheme, we randomly pick a subset from the data and treat points in this subset as the cluster centroids of k-means. These centroids will induce a cluster partition of the whole data, based on which we initialize the component means and covariance matrices likewise, as in the baseline method. Both the second and the third initialization schemes are repeated several times with different random start.

### 3.2 Modal Baum-Welch Algorithm

HMM-VB can be viewed as a special case of a GMM where each component of the GMM corresponds to a particular sequence of states $\mathbf{s} = \{s_1, ..., s_T\}$, that is, a combination of states for all the variable blocks. We call this equivalent GMM the *GMM mapped from HMM-VB*. Each component is a Gaussian distribution with mean $\mu_\mathbf{s} = (\mu_{s_1}^{(1)}, \mu_{s_2}^{(2)}, ..., \mu_{s_T}^{(T)})$ (column-wise stack of vectors) and a covariance matrix, denoted by $\Sigma_\mathbf{s}$, that is block diagonal. The $t$th diagonal block in $\Sigma_\mathbf{s}$ is $\Sigma_{s_t}^{(t)}$ with dimension $d_t \times d_t$. We can thus readily apply the modal clustering framework for GMM to data modeled by HMM-VB. However, as the number of possible sequences grows exponentially with the number of variable blocks $T$, a direct application is computationally infeasible. We discover that because of the block diagonal structure of the covariance matrix of the GMM mapped from HMM-VB, we can in fact avoid computing the posterior of $\mathbf{x}$ belonging to each component (exponentially many of them!). Instead, we only need $L_k(\mathbf{x}, t)$ for all $k$ and $t$ when updating $\mathbf{x}$ in the M-step of MEM. Because the BW algorithm solves $L_k(\mathbf{x}, t)$ at a complexity linear in $T$, we can achieve linear complexity for MEM on HMM-VB as well. We call this new algorithm *Modal Baum-Welch* (MBW)

Algorithm.

For brevity of notation, we use $x^{(t),r}$ to denote the value of the $t$th variable block at iteration $r$. And $\mathbf{x}^{[r]} = (x^{(1),r}, x^{(2),r}, ..., x^{(T),r})$ is the concatenated full vector at iteration $r$. The equivalence of MBW and the Modal EM algorithm is ensured by Theorem 1 below, which is proved in Appendix B.

**Theorem 1.** *For the mapped GMM from a HMM-VB, suppose the solution of the M-step in the MEM algorithm given by Eq. (3) is divided into blocks $\mathbf{x}^{[r+1]} = (x^{(1),r+1}, x^{(2),r+1}, ..., x^{(T),r+1})$. Then*

$$x^{(t),r+1} = \left(\sum_{k \in \mathcal{S}_t} L_k(\mathbf{x}^{[r]}, t) \cdot \left(\Sigma_k^{(t)}\right)^{-1}\right)^{-1} \left(\sum_{k \in \mathcal{S}_t} L_k(\mathbf{x}^{[r]}, t) \cdot \left(\Sigma_k^{(t)}\right)^{-1} \cdot \mu_k^{(t)}\right), \quad t = 1, ..., T.$$

The MBW algorithm iterates the following two steps:

1. E-step: Compute $L_k(\mathbf{x}^{[r]}, t)$, for $k \in \mathcal{S}_t$, $t = 1, ..., T$.

2. M-step:

$$x^{(t),r+1} = \left(\sum_{k \in \mathcal{S}_t} L_k(\mathbf{x}^{[r]}, t) \cdot \left(\Sigma_k^{(t)}\right)^{-1}\right)^{-1} \left(\sum_{k \in \mathcal{S}_t} L_k(\mathbf{x}^{[r]}, t) \cdot \left(\Sigma_k^{(t)}\right)^{-1} \cdot \mu_k^{(t)}\right), \quad t = 1, ..., T.$$

The clustering method based on MBW is straightforward. We first find the state sequence $\mathbf{s}_i^{(*)}$ with maximum posterior given $\mathbf{x}_i$ by the Viterbi algorithm (Young et al. 1997):

$$\mathbf{s}_i^* = \arg\max_{\mathbf{s} \in \hat{\mathcal{S}}} P(\mathbf{s} \mid \mathbf{x}_i), \, i = 1, ..., n.$$

Since different $\mathbf{x}_i$'s may yield the same sequence, we then identify the collection of distinct $\mathbf{s}_i^*$. For each distinct sequence, say $\mathbf{s}_i^*$, find $\mu_{\mathbf{s}_i^*} = (\mu_{s_1^*}^{(1)}, \mu_{s_2^*}^{(2)}, ..., \mu_{s_T^*}^{(T)})$. Use $\mu_{\mathbf{s}_i^*}$ as an initialization for MBW to find the mode associated with it. If $\mu_{\mathbf{s}_i^*}$ and $\mu_{\mathbf{s}_j^*}$ are brought to the same mode by MBW, the data point $\mathbf{x}_i$ and $\mathbf{x}_j$ are put in the same cluster. When $|\hat{\mathcal{S}}|$ is very large, the number of different $\mathbf{s}_i^*$'s can become close to the data size. Hence, the computation we can save by seeking modes starting from $\mu_{\mathbf{s}_i^*}$'s instead of the original data diminishes. As a result, in such cases, we recommend seeking modes directly from the original data. In fact, in the experiment described in Section 4.2, the modes are found in this way.

# 4. Experiments

In this section, we present experiments on several simulated data sets (4.1) and one CyTOF data (4.2). For each data set, the BW algorithm was run repeatedly starting from multiple initial models. In Section 3.1, the different ways of initialization are described. Among the final models, we choose the one yielding the maximum likelihood.

## 4.1 Simulation with Various Types of Variable Block Structures

We first conduct simulation studies to examine the effectiveness of HMM-VB at capturing small clusters and its robustness against different parameter settings.

### 4.1.1. Two variable blocks

Using the same set-up as in Lin et al. (2013), a sample of size $10,000$ with dimension $d = 8$ is drawn from a hierarchical mixture model. There are two variable blocks. Following the notations in the previous section, $\mathbf{x}_i$ is divided into two variable blocks, also called subvectors, $x_i^{(1)}$ and $x_i^{(2)}$. The first subvector contains the first $5$ dimensions: $x_i^{(1)} = (x_{i,1}, ..., x_{i,5})$, with $d_1 = 5$. The second subvector contains the last $3$ dimensions: $x_i^{(2)} = (x_{i,6}, x_{i,7}, x_{i,8})$, with $d_2 = 3$. In particular, $x_i^{(1)}$'s are generated from a mixture of $7$ normal distributions. The last two normal distributions corresponds to rare clusters with component proportions $0.02$ and $0.01$. The two normals have approximately equal mean vectors $(0, 5.5, 5.5, 0, 0)'$, $(0, 6, 6, 0, 0)'$ and common diagonal covariance matrix $2I$. Compared with the last two normal components, the other normal components are designed to have very different mean vectors and larger variances. The last three dimensions, $x_i^{(2)}$'s, are drawn from a mixture of $10$ normal distributions. Specifically, only two of them have high mean values across all three dimensions. The component proportions of $x_i^{(2)}$ vary according to which normal component $x_i^{(1)}$ was generated from, as in the assumption of HMM-VB. The data is designed to have a distinct cluster after standardization (subtract mean and divided by the standard error). In particular, the standardized data have a well-separated region that the five dimensions $x_2$, $x_3$, $x_6$, $x_7$, $x_8$ are of high positive values, and the rest are negative. The subset of interest with size $140$ is indicated in red in Figure 2.

Lin et al. (2013) showed the standard GMM failed completely to identify the subset of interest. Here we contrast the hierarchical Bayesian mixture model approach (Lin et al. 2013) with results from analysis using the new HMM-VB. To fit HMM-VB, we only need to specify the number of mixture components for each variable block, in this example, $M_1$ and $M_2$. If casting as a GMM, the HMM-VB has $M_1 \cdot M_2$ components for the full dimensional data. Model selection using BIC is conducted to select the optimal $M_1$ and $M_2$. Summaries on various model specifications are listed in Table 1. The model with $M_1 = 7$ and $M_2 = 10$ has the lowest BIC, thus it is selected as the optimal model.

Figure 2: Pairwise scatter plots of simulated data described in Section 4.1.1. The subset of interest is plotted in red (140 observations), and the rest are in grey.

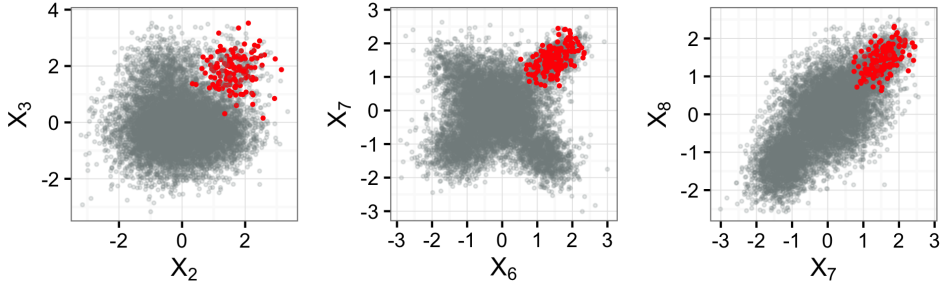

Recall that we search modes by MBW starting from the mean of every component that has been chosen by a data point according to the maximum a posteriori rule. We call the components that have not been chosen by any data point the empty mixture components, while the others the nonempty. Table 1 shows that even though the total number of nonempty mixture components increases with $M_1$ and $M_2$, the number of clusters after modal clustering remains relatively stable. As shown in Figure 3, HMM-VB with $M_1 = 7$ and $M_2 = 10$ correctly identifies 135 observations out of the 140 target sample. It yields similar clustering performance as the hierarchical mixture model developed in Lin et al. (2013), while the hierarchical mixture model can only accommodate two variable blocks.

Table 1: Comparisons of BIC, total number of nonempty mixture components and total number of clusters under various model specifications ($M_1$ and $M_2$) for simulation data in Section 4.1.1.

| $(M_1, M_2)$ | BIC | # nonempty components | # clusters |
| --- | --- | --- | --- |
| (3, 3) | 205,004.0 | 9 | 6 |
| (3, 5) | 201,334.4 | 15 | 7 |
| (7, 10) | 201,045.5 | 67 | 14 |
| (8, 10) | 201,299.3 | 78 | 13 |
| (10, 10) | 201,781.6 | 91 | 15 |
| (10, 15) | 202,482.2 | 138 | 16 |
| (15, 10) | 203,141.6 | 140 | 14 |
| (15, 15) | 203,890.5 | 214 | 18 |
| (20, 15) | 205,300.8 | 252 | 16 |
| (20, 20) | 206,478.6 | 318 | 21 |

Figure 3: Pairwise scatter plots of simulated data from Section 4.1.1. The observations from the subset of interest correctly identified by HMM-VB are plotted in red (135 observations); the observations missed by HMM-VB are plotted in green (5 observations). The observations that are misclassified to the subset of interest by HMM-VB are plotted in blue (26 observations).

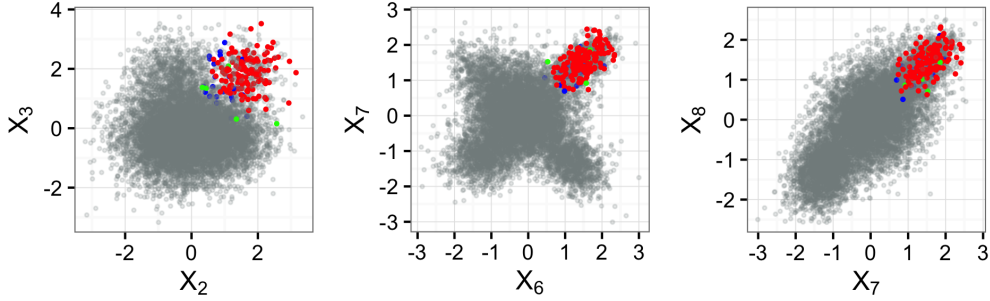

### 4.1.2. Multiple Variable Blocks

A sample of size $10,000$ with $d = 13$ dimensions is drawn such that the first $5$ dimensions are generated from a single multivariate normal distribution with zero mean and diagonal covariance matrix $3I$. The rest $8$ dimensions, divided into 2 variable blocks, are generated according to Section 4.1.1. In summary, the simulated data here have 3 variable blocks with $d_1 = 5, d_2 = 5, d_3 = 3$. Similarly, the data contain a well-separated region such that after data standardization, the five dimensions $x_7, x_8, x_{11}, x_{12}, x_{13}$ are of positive values and the rest are negative. The subset of interest with size 129 is indicated in red in the upper panel of Figure 4.

Table 2 shows that the model with $M_1 = 1$, $M_2 = 6$ and $M_3 = 9$ has the lowest BIC, hence selected as the optimal model for clustering. It should be noted that several model specifications including $(M_1, M_2, M_3) = (1, 7, 10)$ and $(M_1, M_2, M_3) = (1, 6, 9)$ all result in 10 clusters. Hence the simplest model, $(M_1, M_2, M_3) = (1, 6, 9)$, is chosen. As shown in Figure 3, HMM-VB correctly identified 119 observations out of the 129 target sample points.

Next, we compare the clustering result with the generic GMM (see Eq. (1)), which can be taken as a HMM-VB with one variable block. We also apply modal clustering to GMM. Similarly, BIC is computed for models with different numbers of mixture components $M$. Results are listed in Table 3. BIC suggests the $5-$component GMM is the optimal model which identified in total 4 clusters. The subset of interest is completely masked as is shown in Figure 5 middle panel. The $5-$component GMM is unable to identify the correct subset region. We then dramatically increased the number of mixture components to 100. The $100-$component GMM identified in total 69 clusters, significantly larger than the number of clusters determined by the optimal HMM-VB.

Figure 4: Pairwise scatter plots of simulated data from Section 4.1.2. The upper panel contains the pairwise scatter plots on different dimensions. The subset of interest is plotted in red (129 observations). The lower panel contains the same scatter plots as the upper panel, but the subset of interest is instead color coded. The observations from the subset of interest that were correctly identified by HMM-VB are plotted in red (119 observations). The observations that are missed by HMM-VB are plotted in green (10 observations). The observations that are misclassified to the subset of interest by HMM-VB are plotted in blue (13 observations).

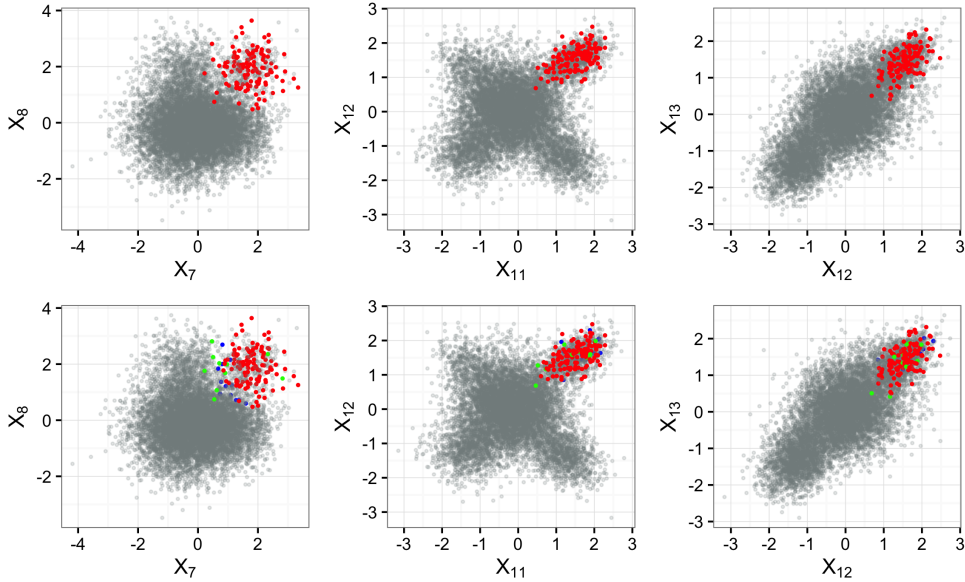

Table 2: Comparisons of BIC, total number of nonempty mixture components and total number of clusters in various model specifications ($M_1$, $M_2$ and $M_3$) for simulation data in Section 4.1.2.

| ($M_1, M_2, M_3$) | BIC | # nonempty components | # clusters |
| --- | --- | --- | --- |
| (1, 3, 5) | 344, 528.4 | 15 | 7 |
| (1, 7, 5) | 343, 873.9 | 34 | 11 |
| (1, 5, 10) | 344, 058.1 | 48 | 8 |
| (1, 6, 9) | 343, 850.6 | 53 | 10 |
| (1, 7, 9) | 344, 217.5 | 57 | 10 |
| (1, 6, 10) | 344, 057.4 | 58 | 10 |
| (2, 5, 5) | 343, 950.6 | 47 | 10 |
| (1, 7, 10) | 344, 312.9 | 67 | 10 |
| (2, 10, 10) | 345, 210.5 | 196 | 12 |
| (10, 10, 10) | 347, 127.6 | 781 | 11 |

Figure 5 bottom panel shows the clustering result for this much over-fitted GMM. Out of the 129 target sample points, 99 observations are correctly identified, whereas HMM-VB correctly identified 119 observations using a much sparser model representation than the $100-$component GMM.

Table 3: Comparisons of BIC and total number of clusters in various model specifications ($M_1$) of GMM for simulation data in Section 4.1.2.

| $M$ | BIC | # clusters |
|---|---|---|
| 4 | $350,280.1$ | 2 |
| 5 | $348,993.2$ | 4 |
| 6 | $349,306.3$ | 4 |
| 8 | $350,062.5$ | 6 |
| 10 | $351,387.2$ | 6 |
| 15 | $355,072.5$ | 7 |
| 20 | $358,906.3$ | 9 |
| 30 | $366,574.4$ | 11 |
| 50 | $381,578.7$ | 17 |
| 100 | $419,425.8$ | 69 |

### 4.1.3. No Information on Variable Blocks

Lastly, we study the clustering performance of HMM-VB when there is no clear-cut variable block structures. A sample of size $10,000$ with $d = 10$ dimensions is drawn from a $50-$component GMM. The component priors for the first 6 components are $\pi_{1:6} = (0.0025, 0.005, 0.1, 0.1, 0.05, 0.03)$, and the other priors are equal, that is, $\pi_s = 0.7125/44 \approx 0.016$, $s = 7, ..., 50$. Furthermore, we make the first two rare components to represent two distinct clusters by setting the first mean vector $\mu_1$ to a vector of 5's and $\mu_2$ to a vector of $-5$'s. The other mean vectors $\mu_{3:50}$ are independently generated from a multivariate normal distribution with 0 mean and identity covariance matrix. The first two covariance matrices $\Sigma_{1:2}$ are both identity matrix. The other covariance matrices $\Sigma_{3:50}$ are independently generated from an inverse Wishart distribution with 15 degrees of freedom and diagonal scale matrix $5I$. Figure 6 shows the scatter plots of the generated data on selected dimensions. The two components are masked by the background data in some dimensions. For example, projected to the 5th and 9th dimensions, the first component (red) cannot be completely separated from the background data (grey).

We first fit the data with HMM-VB using different model specifications. Specifically, we let Model 1 define a HMM-VB with 10 variable blocks such that each variable forms one block. The data is fitted hierarchically from the first variable to the last one, with $M_{1:10} = 10$. Model 2 is defined similarly to Model 1, but we reverse the ordering of the variables such that $x = (x_{10}, x_9, ..., x_1)$. Lastly, we let Model 3 define a HMM-VB with 10 variable blocks such that we randomly reorder $x = (x_{10}, x_9, x_6, x_5, x_4, x_3, x_2, x_1, x_8, x_7)$. Table 4 shows the confusion matrices for the three models respectively. In particular, we assign all the background data that do not belong to the first two components to cluster 0. Table 4 suggests that HMM-VB is quite robust to the ordering of the variables. All the three HMM-VB models can consistently identify the two rare components with high accuracy.

Figure 5: Comparison of clustering performance based on $5-$ and $100-$component GMMs. The upper panel contains the pairwise scatter plots of simulated data from Section 4.1.2 on different selected dimensions. The subset of interest is plotted in red (129 observations). The middle panel contains the same scatter plots as the upper panel, but the observations are color coded according to their cluster membership determined by the $5-$component GMM. The bottom panel contains the same scatter plots as the upper panel, but the subset of interest is color coded. The observations from the subset of interest that were correctly identified by the 100-component GMM are plotted in red (99 observations). The observations that are missed by 100-component GMM are plotted in green (30 observations). The observations that are misclassified to the subset of interest by $100-$component GMM are plotted in blue (14 observations).

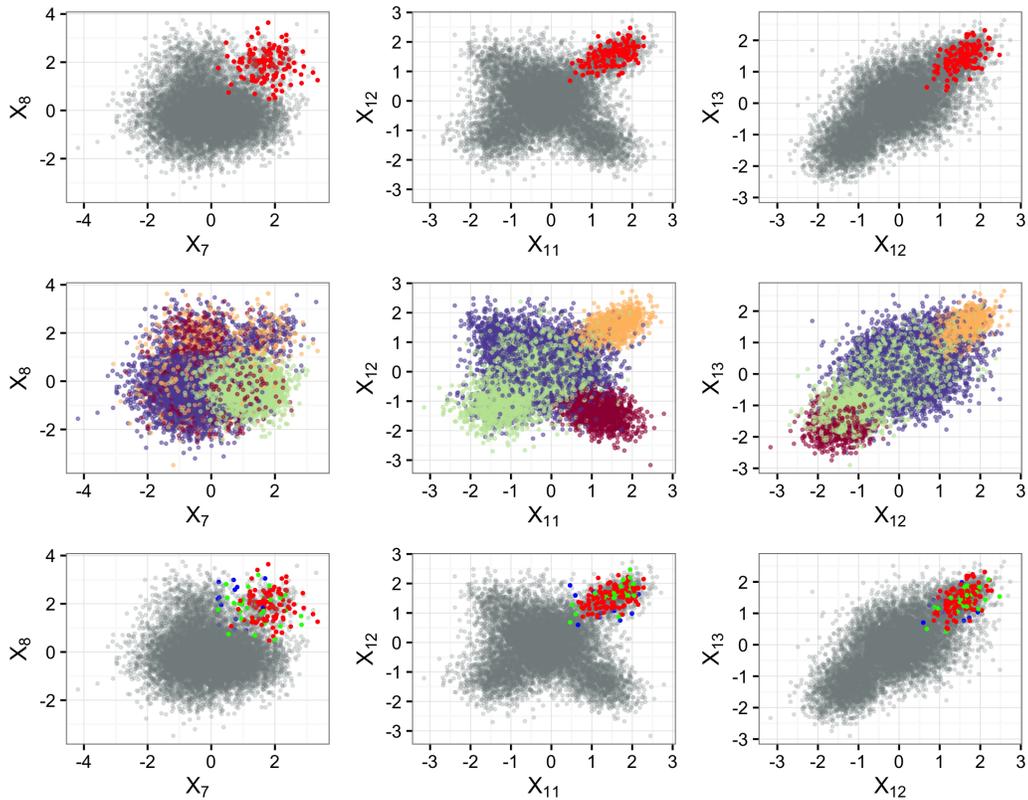

Figure 6: Pairwise scatter plots of simulated data from Section 4.1.3. The two subsets of interest are plotted in red for component 1 (25 observations) and blue for component 2 (52 observations).

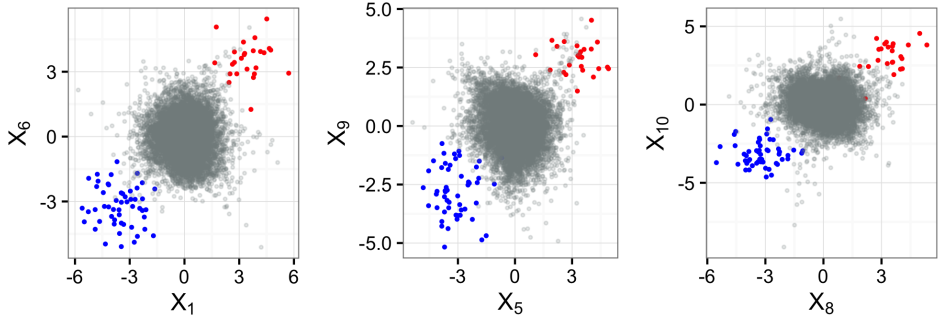

Table 4: Comparison of clustering performance based on three HMM-VB models for simulation data in Section 4.1.3.

|  |  | True | | |
|---|---|---|---|---|
|  |  | 0 | 1 | 2 |
|  | 0 | $9,923$ | 3 | 0 |
| Model 1 | 1 | 0 | 22 | 0 |
|  | 2 | 0 | 0 | 52 |
|  | 0 | $9,923$ | 3 | 0 |
| Model 2 | 1 | 0 | 22 | 0 |
|  | 2 | 0 | 0 | 52 |
|  | 0 | $9,923$ | 2 | 2 |
| Model 3 | 1 | 0 | 23 | 0 |
|  | 2 | 0 | 0 | 50 |

We then contrast the above results with GMM. GMM with $M = 50$ is first fitted. It breaks the first and second rare components into 3 and 7 distinct clusters respectively. Although $M = 50$ is the right specification for the model, the estimation is not sufficiently accurate to capture the rare clusters. Next, $M$ is reduced to 20. This $20-$component GMM can accurately detect the first component with 100% accuracy. However, the second component is divided into 3 groups of sizes 17, 20, and 15, each belonging to a distinct cluster. When the number of components $M$ is further reduced to 10, the second component is again broken up into 2 distinct clusters, while the first component is completely masked by the background data.

We see that even in the lack of the variable block structure, HMM-VB can still outperform

GMM in identifying multiple rare clusters. This result suggests that HMM-VB is a strategy to form parsimonious versions of GMM which can be more effective than the common way of varying the number of components.

## 4.2 Study of CyTOF Data

In this section, we study the performance of HMM-VB on a higher dimensional data set obtained from CyTOF experiment (Becher et al. 2014). The particular data set that we analyze here is from mouse lung sample obtained from three C57BI6 wild-type mice and three Csf2rb$^{-/-}$ mice, which in total contains $46,204$ single cells with $39$ measured cell markers. According to the gating hierarchy provided in Becher et al. (2014), it defines roughly 11 variable blocks, with maximum block size $8$ and minimum block size 1. Becher et al. (2014) performed automated clustering on a projected (latent) $2-$dimensional space by first using nonlinear dimension reduction technique on the original $39-$dimensional space. However, it has been studied, e.g. Lin et al. (2015b), that dimension reduction generates a "cluttered" display, and this can prevent density estimation from accurately representing low probability regions.

We first standardize the data. The reason is that the moderately high-dimensional data has a nearly singular covariance matrix (that is, when a single Gaussian is fit). This prevents the direct use of GMM. It should be noted that HMM-VB encounters no difficulty in fitting the data because of the smaller dimensions of the individual variable blocks. In order to compare GMM and HMM-VB, we work on the standardized data through the rest of the section.

We fit the data using HMM-VB with specification as follows. There are 11 variable blocks. For the variable blocks $i$ with dimension lower than $4$, we set the corresponding number of mixture components $M_i = 5$. For variable blocks $j$ with dimension between $5$ and 7, we set the corresponding number of mixture components $M_j = 10$. For the other variable blocks, we set the corresponding number of mixture components to $15$. After modal clustering, this model results in $825$ clusters. The average CPU time per model fitting is $11.4$ min on an iMac with Intel i7 Processor at $3.0$ GHz with 8GB memory. We next fit the data using GMM with $M = 500$. On the same iMac, the model fitting takes $284.2$ min, which is $25$ times longer than did HMM-VB. After modal clustering, it results in $316$ clusters.

We compare the performance of the two models for clustering a relatively low probability region ($6\%$ of total cells), which is shown in Figure 7. Specifically, Figure 7 compares the finer cellular compositions of one well separated region, which is visualized on two latent dimensions obtained from a dimension reduction technique applied to the original 39 dimensions, as provided by the result of Becher et al. (2014). Becher et al. (2014) also studied the effectiveness of such dimension reduction result in showing clusters of major cell types. HMM-VB (right subplot) uses $18$ clusters to define this particular region. GMM (left subplot) uses $40$ clusters to define the same region. However, some of the clusters include cells that are visually far away from the particular region. This indicates that GMM has difficulty in estimating the structure of this moderately high-dimensional data.

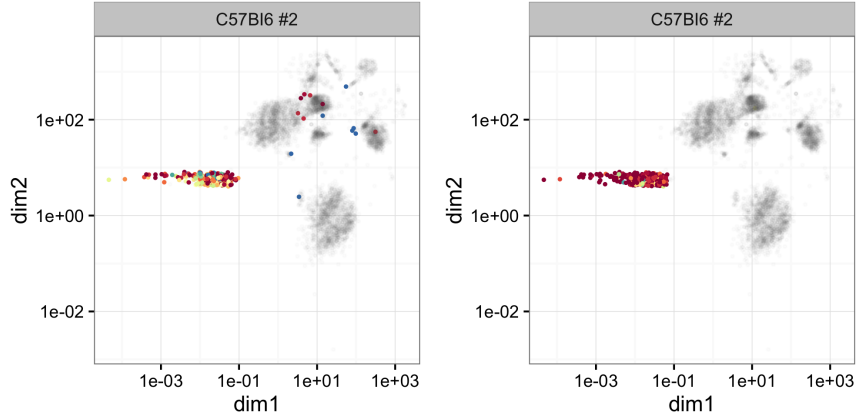

Figure 7: Left: GMM clustering analysis for a particular data region of one selected mouse. Right: The corresponding HMM-VB clustering analysis. The data is shown in grey. Clustering results are in different colors with one color defines one cluster.

## 5. Discussion

We develop and explore a novel hierarchical mixture model (HMM-VB) with the goal of automated clustering of large-scale data sets that contain multiple rare clusters. The method exploits the natural structure among groups of variables for more effective modeling and clustering analysis. Both the simulated and real data examples demonstrate the effectiveness of HMM-VB for identifying multiple rare clusters in large-scale data sets. The practical motivation is from the automated analysis of single-cell cytometry data. The standard visual gating approach suffers severe limitations including non-reproducibility and labor intensiveness. The cytometry field is increasingly interested in using more consistent automated statistical methods. Standard mixture models, however, are not powerful enough to detect clusters of very low probabilities when applied to large data sets. One key feature of the new model is by design the ability to quantify the inherent chain-like dependence among groups of variables for more effective clustering, especially in identifying rare clusters that are almost undetectable by existing mixture modeling approaches.

Technically, our clustering method integrates two new algorithms, one for fitting HMM-VB and the other for performing modal clustering, both necessary to make HMM-VB a practical tool for large-scale data analysis. In theory, the number of components grows exponentially with the num-ber of variable groups. Existing methods for estimation and mode identification thus bear exponential complexity, rendering them infeasible on real-world data. We derive and implement algorithms with linear complexity in the number of variable blocks for both tasks.

The development of more structured mixture models for clustering is necessary for large-scale data, especially single-cell data as the technologies advance and the number of markers that can be measured increases. However, due to the huge number of mixture components, many tiny clusters can be generated. The total number of points in those clusters can be a small fraction of the entire data. For instance, nearly half of the clusters identified in the analysis of CyTOF data are of size one. In this situation, more sophisticated methods that measure the separability of these tiny clusters and other larger ones can be used for modal clustering (e.g., Lee and Li (2012)). The development of HMM-VB can also be used to determine the optimal variable groupings and orderings based on some model selection criterion, such as BIC. This can be useful for developing new cytometry experiment for providing objective optimal gating hierarchy.

# Appendix A

We present the forward-backward algorithm for computing $L_k(t)$ and $H_{k,l}(t)$ efficiently. Define the forward probability $\alpha_k(t)$ as the joint probability of observing the first $t$ vectors $x_\tau$, $\tau = 1, ..., t$, and being in state $k$ at time $t$:

$$\alpha_k(t) = P(x_1, x_2, ..., x_t, s_t = k).$$

This probability can be evaluated by the following recursive formula:

$$\begin{aligned} \alpha_k(1) &= \pi_k b_k(x_1), \quad 1 \leq k \leq M, \\ \alpha_k(t) &= b_k(x_t) \sum_{l=1}^{M} \alpha_l(t-1) a_{l,k}, \quad 1 < t \leq T,\ 1 \leq k \leq M. \end{aligned}$$

Define the backward probability $\beta_k(t)$ as the conditional probability of observing the vectors after time $t$, $x_\tau$, $\tau = t+1, ..., T$, given the state at time $t$ is $k$.

$$\begin{aligned} \beta_k(t) &= P(x_{t+1}, ..., x_T \mid s_t = k), \quad 1 \leq t \leq T-1, \\ \beta_k(T) &= 1, \quad \text{for all } k. \end{aligned}$$

As with the forward probability, the backward probability can be evaluated using the following

recursion:

$$\begin{aligned}
\beta_k(T) &= 1, \\
\beta_k(t) &= \sum_{l=1}^{M} a_{k,l} b_l(x_{t+1}) \beta_l(t+1), \quad 1 \leq t < T.
\end{aligned}$$

The probabilities $L_k(t)$ and $H_{k,l}(t)$ are solved by

$$L_k(t) = P(s_t = k \mid \mathbf{x}) = \frac{P(\mathbf{x}, s_t = k)}{P(\mathbf{x})} = \frac{1}{P(\mathbf{x})} \alpha_k(t) \beta_k(t),$$

$$\begin{aligned}
H_{k,l}(t) &= P(s_t = k, s_{t+1} = l \mid \mathbf{u}) = \frac{P(\mathbf{x}, s_t = k, s_{t+1} = l)}{P(\mathbf{x})} \\
&= \frac{1}{P(\mathbf{x})} \alpha_k(t) a_{k,l} b_l(x_{t+1}) \beta_l(t+1),
\end{aligned}$$

where $P(\mathbf{x}) = \sum_{k=1}^{M} \alpha_k(t) \beta_k(t)$.

We now provide the Baum-Welch algorithm for the case of multiple sequences. For brevity, assume all the sequences are of length $T$. Denote the $i$th sequence by $\mathbf{x}_i = \{x_{i,1}, x_{i,2}, ..., x_{i,T}\}$, $i = 1, ..., n$. In each iteration, we compute the forward and backward probabilities for each sequence separately in the same way as previously described. We also compute $L_k(t)$ and $H_{k,l}(t)$ separately for each sequence. As a general pattern of notations, we put a superscript $(i)$ to indicate the quantities for the $i$th sequence.

1. Compute the forward and backward probabilities $\alpha_k^{(i)}(t)$, $\beta_k^{(i)}(t)$, $k = 1, ..., M$, $t = 1, ..., T$, $i = 1, ..., n$, under the current set of parameters.

$$\begin{aligned}
\alpha_k^{(i)}(1) &= \pi_k b_k(x_{i,1}), \quad 1 \leq k \leq M,\ 1 \leq i \leq n, \\
\alpha_k^{(i)}(t) &= b_k(x_{i,t}) \sum_{l=1}^{M} \alpha_l^{(i)}(t-1) a_{l,k}, \quad 1 < t \leq T,\ 1 \leq k \leq M,\ 1 \leq i \leq n, \\
\beta_k^{(i)}(T) &= 1, \quad 1 \leq k \leq M,\ 1 \leq i \leq n, \\
\beta_k^{(i)}(t) &= \sum_{l=1}^{M} a_{k,l} b_l(x_{i,t+1}) \beta_l^{(i)}(t+1), \quad 1 \leq t < T,\ 1 \leq k \leq M,\ 1 \leq i \leq n.
\end{aligned}$$

2. Compute $L_k^{(i)}(t)$, $H_{k,l}^{(i)}(t)$ using $\alpha_k^{(i)}(t)$, $\beta_k^{(i)}(t)$. Let $P(\mathbf{x}_i) = \sum_{k=1}^{M} \alpha_k^{(i)}(1)\beta_k^{(i)}(1)$.

$$L_k^{(i)}(t) = \frac{1}{P(\mathbf{x}_i)} \alpha_k^{(i)}(t)\beta_k^{(i)}(t),$$

$$H_{k,l}^{(i)}(t) = \frac{1}{P(\mathbf{x}_i)} \alpha_k^{(i)}(t) a_{k,l} b_l(x_{i,t+1}) \beta_l^{(i)}(t+1).$$

3. Update the parameters using $L_k^{(i)}(t)$, $H_{k,l}^{(i)}(t)$.

$$\mu_k = \frac{\sum_{i=1}^{n} \sum_{t=1}^{T} L_k^{(i)}(t) x_{i,t}}{\sum_{i=1}^{n} \sum_{t=1}^{T} L_k^{(i)}(t)},$$

$$\Sigma_k = \frac{\sum_{i=1}^{n} \sum_{t=1}^{T} L_k^{(i)}(t)(x_{i,t} - \mu_k)(x_{i,t} - \mu_k)'}{\sum_{i=1}^{n} \sum_{t=1}^{T} L_k^{(i)}(t)},$$

$$a_{k,l} = \frac{\sum_{i=1}^{n} \sum_{t=1}^{T-1} H_{k,l}^{(i)}(t)}{\sum_{i=1}^{n} \sum_{t=1}^{T-1} L_k^{(i)}(t)}.$$

## Appendix B

We prove the equivalence of MBM and MEM for HMM-VB. It is clear that the density of HMM-VB in Eq. (6) is a mixture model if we take the state sequence $\mathbf{s} \in \hat{\mathcal{S}}$ as the index for the mixture component. Each component is a Gaussian distribution with mean $\mu_\mathbf{s} = (\mu_{s_1}^{(1)}, \mu_{s_2}^{(2)}, ..., \mu_{s_T}^{(T)})'$ (column-wise stack of vectors) and a covariance matrix, denoted by $\Sigma_\mathbf{s}$, that is block diagonal. The $t$th diagonal block in $\Sigma_\mathbf{s}$ is $\Sigma_{s_t}^{(t)}$ with dimension $d_t \times d_t$.

$$\Sigma_\mathbf{s} = \begin{pmatrix} \Sigma_{s_1}^{(1)} & 0 & 0 & \cdots & 0 \\ 0 & \Sigma_{s_2}^{(2)} & 0 & \cdots & 0 \\ \cdots & \cdots & \cdots & \cdots & \cdots \\ 0 & 0 & 0 & \cdots & \Sigma_{s_T}^{(T)} \end{pmatrix}.$$

If we apply MEM directly to HMM-VB and keep in mind that $\mathbf{s}$ is the index for the mixture component, we need to compute the posterior $P(\mathbf{s} \mid \mathbf{x})$ in the E-step and

$$\left( \sum_{\mathbf{s} \in \hat{\mathcal{S}}} P(\mathbf{s} \mid \mathbf{x}) \Sigma_\mathbf{s}^{-1} \right)^{-1} \left( \sum_{\mathbf{s} \in \hat{\mathcal{S}}} P(\mathbf{s} \mid \mathbf{x}) \Sigma_\mathbf{s}^{-1} \mu_\mathbf{s} \right)$$

in the M-step. The computational hurdle is that the number of possible sequences $\mathbf{s}$, that is, $|\hat{\mathcal{S}}|$, grows exponentially with $T$ (assuming similar $|\mathcal{S}_t|$'s).

Because $\Sigma_\mathbf{s}$ is block diagonal, we have

$$\left(\sum_{\mathbf{s}\in\hat{\mathcal{S}}} P(\mathbf{s}\mid \mathbf{x})\Sigma_\mathbf{s}^{-1}\right)^{-1} \left(\sum_{\mathbf{s}\in\hat{\mathcal{S}}} P(\mathbf{s}\mid \mathbf{x})\Sigma_\mathbf{s}^{-1}\mu_\mathbf{s}\right)$$

$$= \begin{pmatrix} \left(\sum_{\mathbf{s}\in\hat{\mathcal{S}}} P(\mathbf{s}\mid \mathbf{x})\left(\Sigma_{s_1}^{(1)}\right)^{-1}\right)^{-1} \left(\sum_{\mathbf{s}\in\hat{\mathcal{S}}} P(\mathbf{s}\mid \mathbf{x})\left(\Sigma_{s_1}^{(1)}\right)^{-1}\mu_{s_1}^{(1)}\right) \\ \left(\sum_{\mathbf{s}\in\hat{\mathcal{S}}} P(\mathbf{s}\mid \mathbf{x})\left(\Sigma_{s_2}^{(2)}\right)^{-1}\right)^{-1} \left(\sum_{\mathbf{s}\in\hat{\mathcal{S}}} P(\mathbf{s}\mid \mathbf{x})\left(\Sigma_{s_2}^{(2)}\right)^{-1}\mu_{s_2}^{(2)}\right) \\ \vdots \\ \vdots \\ \left(\sum_{\mathbf{s}\in\hat{\mathcal{S}}} P(\mathbf{s}\mid \mathbf{x})\left(\Sigma_{s_T}^{(T)}\right)^{-1}\right)^{-1} \left(\sum_{\mathbf{s}\in\hat{\mathcal{S}}} P(\mathbf{s}\mid \mathbf{x})\left(\Sigma_{s_T}^{(T)}\right)^{-1}\mu_{s_T}^{(T)}\right) \end{pmatrix}$$

Hence the $t$th variable block of $\mathbf{x}$ is given by

$$x^{(t)} = \left(\sum_{\mathbf{s}\in\hat{\mathcal{S}}} P(\mathbf{s}\mid \mathbf{x})\left(\Sigma_{s_t}^{(t)}\right)^{-1}\right)^{-1} \left(\sum_{\mathbf{s}\in\hat{\mathcal{S}}} P(\mathbf{s}\mid \mathbf{x})\left(\Sigma_{s_t}^{(t)}\right)^{-1}\mu_{s_t}^{(t)}\right), \quad t=1,2,...,T.$$

Let $I(\cdot)$ be the indicator function that equals 1 when the argument is true. Note that

$$\sum_{\mathbf{s}\in\hat{\mathcal{S}}} P(\mathbf{s}\mid \mathbf{x})\left(\Sigma_{s_t}^{(t)}\right)^{-1} = \sum_{k\in\mathcal{S}_t}\left[\sum_{\mathbf{s}\in\hat{\mathcal{S}}} P(\mathbf{s}\mid \mathbf{x})I(s_t=k)\right]\cdot \left(\Sigma_k^{(t)}\right)^{-1}$$

$$= \sum_{k\in\mathcal{S}_t} L_k(\mathbf{x},t)\cdot \left(\Sigma_k^{(t)}\right)^{-1}$$

according to the definition of $L_k(\mathbf{x},t)$ in Eq. (8). By the same technique, we can show that

$$\sum_{\mathbf{s}\in\hat{\mathcal{S}}} P(\mathbf{s}\mid \mathbf{x})\left(\Sigma_{s_t}^{(t)}\right)^{-1}\mu_{s_t}^{(t)} = \sum_{k\in\mathcal{S}_t} L_k(\mathbf{x},t)\cdot \left(\Sigma_k^{(t)}\right)^{-1}\mu_k^{(t)}.$$

Thus we have proved the equivalence of MBW and MEM for HMM-VB.

# References

Aghaeepour, N., Finak, G., Hoos, H., Mosmann, T. R., Brinkman, R., Gottardo, R., and Scheuermann, R. H. (2013). Critical assessment of automated flow cytometry data analysis techniques.

*Nature Methods*, 10(3):228–238. 3

Aghaeepour, N., Nikolic, R., Hoos, H. H., and Brinkman, R. R. (2011). Rapid cell population identification in flow cytometry data. *Cytometry Part A*, 79A(1):6–13. 3

Bandura, D. R., Baranov, V. I., Ornatsky, O. I., Antonov, A., Kinach, R., Lou, X., Pavlov, S., Vorobiev, S., Dick, J. E., and Tanner, S. D. (2009). Mass cytometry: Technique for real time single cell multitarget immunoassay based on inductively coupled plasma time-of-flight mass spectrometry. *Analytical Chemistry*, 81(16):6813–6822. PMID: 19601617. 1

Becher, B., Schlitzer, A., Chen, J., Mair, F., Sumatoh, H. R., Teng, K. W. W., Low, D., Ruedl, C., Riccardi-Castagnoli, P., Poidinger, M., Greter, M., Ginhoux, F., and Newell, E. W. (2014). High-dimensional analysis of the murine myeloid cell system. *Nature Immunology*, 15(12):1181–1189. Resource. 22

Boedigheimer, M. J. and Ferbas, J. (2008). Mixture modeling approach to flow cytometry data. *Cytometry Part A*, 73A(5):421–429. 3

Celeux, G., Hurn, M., and Robert, C. P. (2000). Computational and inferential difficulties with mixture posterior distributions. *Journal of the American Statistical Association*, 95(451):957–970. 4

Chan, C., Feng, F., Ottinger, J., Foster, D., West, M., and Kepler, T. B. (2008). Statistical mixture modeling for cell subtype identification in flow cytometry. *Cytometry Part A*, 73A(8):693–701. 3

Chan, C., Lin, L., Frelinger, J., Hebert, V., Gagnon, D., Landry, C., Sékaly, R.-P., Enzor, J., Staats, J., Weinhold, K. J., Jaimes, M., and West, M. (2010). Optimization of a highly standardized carboxyfluorescein succinimidyl ester flow cytometry panel and gating strategy design using discriminative information measure evaluation. *Cytometry Part A*, 77A(12):1126–1136. 3

Chattopadhyay, P. K., Gierahn, T. M., Roederer, M., and Love, J. C. (2014). Single-cell technologies for monitoring immune systems. *Nature Immunology*, 15(2):128–135. Review. 1

Ciuffreda, D., Comte, D., Cavassini, M., Giostra, E., Bühler, L., Perruchoud, M., Heim, M. H., Battegay, M., Genné, D., Mulhaupt, B., Malinverni, R., Oneta, C., Bernasconi, E., Monnat, M., Cerny, A., Chuard, C., Borovicka, J., Mentha, G., Pascual, M., Gonvers, J.-J., Pantaleo, G., and Dutoit, V. (2008). Polyfunctional hcv-specific t-cell responses are associated with effective control of hcv replication. *European Journal of Immunology*, 38(10):2665–2677. 1

Corey, L., Gilbert, P. B., Tomaras, G. D., Haynes, B. F., Pantaleo, G., and Fauci, A. S. (2015). Immune correlates of vaccine protection against hiv-1 acquisition. *Science Translational Medicine*, 7(310):310rv7–310rv7. 1

Dempster, A. P., Laird, N. M., and Rubin, D. B. (1977). Maximum likelihood from incomplete data via the em algorithm. *Journal of the Royal Statistical Society: Series B(Statistical Methodology)*, 39(1):1–38. 6

Finak, G., Bashashati, A., Brinkman, R., and Gottardo, R. (2009). Merging mixture components for cell population identification in flow cytometry. *Advances in Bioinformatics*, Article ID 247646. 3

Lee, H. and Li, J. (2012). Variable selection for clustering by separability based on ridgelines. *Journal of Computational and Graphical Statistics*, 21(2):315–337. 7, 24

Li, J., Ray, S., and Lindsay, B. G. (2007). A nonparametric statistical approach to clustering via mode identification. *Journal of Machine Learning Research*, 8(8):1687–1723. 3, 5, 6, 7

Lin, L., Chan, C., Hadrup, S. R., Froesig, T. M., Wang, Q., and West, M. (2013). Hierarchical bayesian mixture modelling for antigen-specific t-cell subtyping in combinatorially encoded flow cytometry studies. *Statistical Applications in Genetics and Molecular Biology*, 12(3):309–331. 4, 15, 16

Lin, L., Chan, C., and West, M. (2016). Discriminative variable subsets in bayesian classification with mixture models, with application in flow cytometry studies. *Biostatistics*, 17(1):40–53. 3

Lin, L., Finak, G., Ushey, K., Seshadri, C., Hawn, T. R., Frahm, N., Scriba, T. J., Mahomed, H., Hanekom, W., Bart, P.-A., Pantaleo, G., Tomaras, G. D., Rerks-Ngarm, S., Kaewkungwal, J., Nitayaphan, S., Pitisuttithum, P., Michael, N. L., Kim, J. H., Robb, M. L., O'Connell, R. J., Karasavvas, N., Gilbert, P., C De Rosa, S., McElrath, M. J., and Gottardo, R. (2015a). COMPASS identifies t-cell subsets correlated with clinical outcomes. *Nature Biotechnology*, 33(6):610–616. 1

Lin, L., Frelinger, J., Jiang, W., Finak, G., Seshadri, C., Bart, P.-A., Pantaleo, G., McElrath, J., DeRosa, S., and Gottardo, R. (2015b). Identification and visualization of multidimensional antigen-specific t-cell populations in polychromatic cytometry data. *Cytometry Part A*, 87(7):675–682. 22

Lo, K., Brinkman, R. R., and Gottardo, R. (2008). Automated gating of flow cytometry data via robust model-based clustering. *Cytometry Part A*, 73A(4):321–332. 3

Maecker, H. T., McCoy, J. P., and Nussenblatt, R. (2012). Standardizing immunophenotyping for the human immunology project. *Nature Reviews Immunology*, 12(3):191–200. 1

Melnykov, V. and Maitra, R. (2010). Finite mixture models and model-based clustering. *Statistics Surveys*, 4:80–116. 6

Perfetto, S. P., Chattopadhyay, P. K., and Roederer, M. (2004). Seventeen-colour flow cytometry: unravelling the immune system. *Nature Reviews Immunology*, 4(8):648–655. 1

Pyne, S., Hu, X., Wang, K., Rossin, E., Lin, T.-I., Maier, L. M., Baecher-Allan, C., McLachlan, G. J., Tamayo, P., Hafler, D. A., De Jager, P. L., and Mesirov, J. P. (2009). Automated high-dimensional flow cytometric data analysis. *Proceedings of the National Academy of Sciences*, 106(21):8519–8524. 3

Ray, S. and Pyne, S. (2012). A computational framework to emulate the human perspective in flow cytometric data analysis. *PloS one*, 7(5):e35693. 8

Richardson, S. and Green, P. J. (1997). On bayesian analysis of mixtures with an unknown number of components (with discussion). *Journal of the Royal Statistical Society: Series B (Statistical Methodology)*, 59(4):731–792. 4

Seshadri, C., Lin, L., Scriba, T. J., Peterson, G., Freidrich, D., Frahm, N., DeRosa, S. C., Moody, D. B., Prandi, J., Gilleron, M., Mahomed, H., Jiang, W., Finak, G., Hanekom, W. A., Gottardo, R., McElrath, M. J., and Hawn, T. R. (2015). T cell responses against mycobacterial lipids and proteins are poorly correlated in south african adolescents. *The Journal of Immunology*, 195(10):4595–4603. 1

Spitzer, M. H. and Nolan, G. P. (2016). Mass cytometry: Single cells, many features. *Cell*, 165(4):780–791. 1

Stephens, M. (2000). Dealing with label switching in mixture models. *Journal of the Royal Statistical Society: Series B (Statistical Methodology)*, 62(4):795–809. 4

Young, S., Evermann, G., Gales, M., Hain, T., Kershaw, D., Liu, X., Moore, G., Odell, J., Ollason, D., Povey, D., Valtchev, V., and Woodland, P. (1997). *The HTK Book*. Cambridge University Press. 9, 14